# Transformation bending device emulated by graded-index waveguide

Y. Wang, C. Sheng, H. Liu,[*] Y. J. Zheng, C. Zhu, S. M. Wang, and S. N. Zhu

*Department of Physics, National Laboratory of Solid State Microstructures, Nanjing University, Nanjing 210093, China*
[*]*liuhui@nju.edu.cn*;
URL: *http://dsl.nju.edu.cn/mpp*

**Abstract:** We demonstrate that a transformation device can be emulated using a gradient-index waveguide. The effective index of the waveguide is spatially varied by tailoring a gradient thickness dielectric waveguide. Based on this technology, we demonstrate a transformation device guiding visible light around a sharp corner, with low scattering loss and reflection loss. The experimental results are in good agreement with the numerical results.

## 1. Introduction

Since the pioneering theoretical work of Pendry and Leonhardt, the duo that designed the well-known spherical cloaks and invisibility devices capable of guiding electromagnetic waves around embedded obstacles [1, 2], researchers in the field of optics have been greatly attracted to the possibility of precisely controlling the flow of light by modifying the microscopic response of an artificial optical material based on principles of transformation optics. With the aid of metamaterials whose effective permittivity and permeability can be manipulated, the experimental demonstration of cloaking has been achieved at microwave frequencies using split-ring resonator metamaterials [3]. The appeal of transformation optics, however, goes beyond invisibility: it can create fascinating effects, such as electromagnetic wormholes [4], hidden gateways [5], and optical black holes [6, 7]. A wide variety of conventional devices can be redesigned with the application of transformation optics, such as field concentrators [8], field rotators [9, 10], field shifters [11], bending waveguides [12-15], as well as many types of lens [16, 17] and other advanced devices [18, 19]. However, at optical frequencies, metamaterials with very complicated permittivity and permeability are usually accompanied by intolerable ohmic losses and scattering losses that hinder the usefulness of such devices. As a result, significant effort has been put into graded dielectric nanostructures as a means to avoid ohmic loss at optical frequencies [20-23]. Nanostructures, being inhomogeneous devices, contain abrupt discontinuities such as holes or different layers that cause scattering losses. Gradient-index (GRIN) optics [24], by comparison, significantly reduces scattering losses. In fact, based on the gradual change of the effective index, more sophisticated elements like the Maxwell fish-eye lens, the Luneburg lens, and the Eaton lens have been proposed more than half a century ago [25]. Recently, with the aid of tapered waveguides [26, 27] that can gradually change effective index, the optical cloaking and super-resolution imaging used to require anisotropic dielectric permittivity and magnetic permeability can be emulated with low-loss broadband performance in the visible frequency range. Therefore, tapered graded-index waveguides are a better choice to realize transformation optics devices in the visible region.

In this paper, we demonstrate that a transformation bending device can be experimentally emulated using a homogeneous, graded dielectric waveguide. The guiding layer of the waveguide is a dielectric film of graded effective refractive index owing to the slow-changing thickness. The effective index of this configuration varies gradually in a truly continuous manner, which results in low-loss broadband performance in the visual frequency range. This is an achievement that is difficult to achieve by other means. We apply this technology to transport light around a 90 ° corner without reflection. Through the fluorescence image of QDs doped in the dielectric layer being excited by light transported along the waveguide, we directly observe the light propagating around the corner without reflection or scattering. A waveguide theory model is given to analyze such system. The simulated results are in good agreement with the experiments.

The paper is composed of six parts. Section 2 discusses the fabrication process and surface morphology of the structure. Theoretical analysis and numerical simulation are discussed in Section 3, followed by optical measurement results in Section 4. Finally, the discussion and outlook are presented in Section 5 and the conclusion in Section 6.

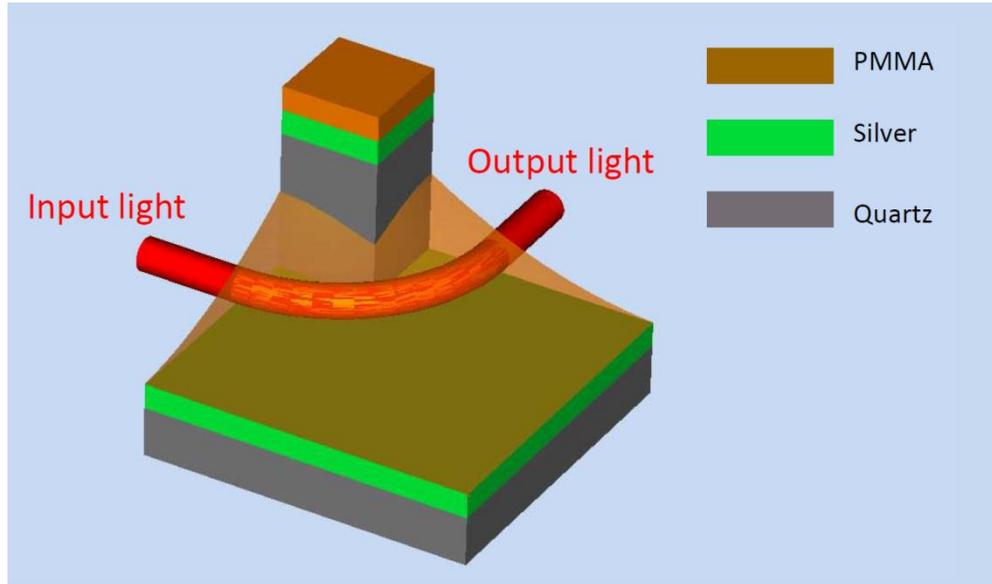

Fig. 1. Schematic picture of graded-index waveguide around a right-angled corner.

## 2. Device fabrication and morphological characterization

The gradient index waveguide designed for this study is depicted in Fig. 1. A gradient thickness poly(methyl methacylate) (PMMA) film is embedded around a sharp quartz corner, between open air and 50 nm silver film supported by a quartz substrate. The thickness is homogeneous in its tangential direction and tapers off along its radial direction, defining the top corner as the origin of the refractive index according to the waveguide theory. When passing through the structure, light approaches the high index area and makes a right-angle turn.

The sample fabrication process is based on the spin-coating technique shown on Fig. 2. First, two quartz wafers are carefully glued together to form a 90 ° corner (Figs. 2(a) and (b)). Then, a 50-nm thick silver film is sputtered onto the substrate (Fig. 2(c)). Some gratings are milled near the corner on the silver film using the focused ion beam (FIB) technique. Afterward, a layer of PMMA containing CdSe/ZnS quantum dots is spin-coated and baked at 70 ℃ for two hours. Due to surface tension, some PMMA adsorb on the corner and self-assemble into a gradient slope (Fig. 2(d)). As a result, a ditch is left near the slope. Finally, a graded waveguide can be obtained at the PMMA slope.

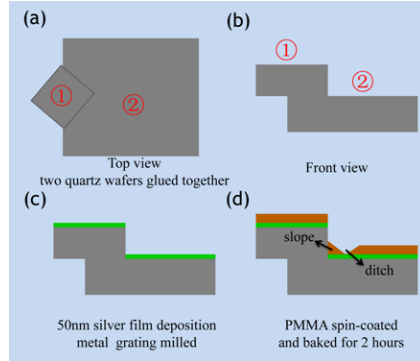

Fig. 2. Structure fabrication (a) – (d). Schematic diagram of the fabrication process.

Figure. 3(a) is the microscopy image of the sample illuminated by a white light source. The blurred interference pattern is caused by the thickness change and indicates a gradual thickness variation around the corner. A coherent 633 nm laser is used to illuminate the sample to obtain a clear interference pattern to measure the surface profile. The red square area of 80 microns in Fig. 3(a) is magnified using a higher-magnification objective lens. A clear interference stripe is shown in Fig. 3(b), which is caused by the PMMA's graded thickness. We use FIB to drill a rectangular hole near the corner to directly observe the PMMA's change in thickness; its cross-section view is shown in Fig. 3(c) (note that the picture is taken at a 45-degree angle to the sample plane and the corner is to the left of the hole). We can see that the thickness of the PMMA layer (indicated by the gray shadow region) does indeed taper off as it moves away from the corner and approaches zero. Using interference theory, we can calculate the thickness of PMMA based on the interference pattern given in Fig. 3(b). The calculated profile of thickness distribution around the corner is given in Fig. 3(d). The results show that the thickness gradually decreases from about 1.5 microns at the corner to zero as distance from the corner increases.

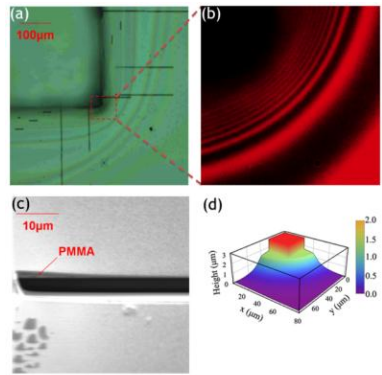

Fig. 3. Microscopy interference pattern of the sample illuminated by white light (a) and by a laser at 633 nm (b); (c) FIB image of a cross-section near the sharp corner; (d) thickness profile around the corner retrieved from (b) and (c).

## 3. Theoretical analysis and numerical simulation

Based on the measured thickness profile, the distribution of index around the corner can be calculated according to the following waveguide theory. For the structure given in Fig. 4(a), the PMMA layer over metal film can be seen as a dielectric waveguide. Considering that the polarization of the incident laser used in our experiment is perpendicular to the coupling gratings, only TM mode can be excited. Generally, the effective index of its waveguide modes is determined by the thickness of the PMMA layer. This result can be analytically deduced from the dispersion equation (TM mode) as

$$\exp[2 \cdot i \cdot (\theta_{12} - \theta_{23} - \kappa_2 \cdot d)] = \frac{\exp[2 \cdot \gamma_3 \cdot t + 2 \cdot i \cdot (\theta_{34} - \theta_{23})] + 1}{\exp[2 \cdot \gamma_3 \cdot t + 2 \cdot i \cdot (\theta_{34} + \theta_{23})] + 1},$$

where $\theta_{12} = \arctan\left(\frac{\gamma_1 \cdot \varepsilon_2}{\kappa_2 \cdot \varepsilon_1}\right)$, $\theta_{23} = \arctan\left(\frac{\gamma_3 \cdot \varepsilon_2}{\kappa_2 \cdot \varepsilon_3}\right)$, $\theta_{34} = \arctan\left(\frac{\kappa_4 \cdot \varepsilon_3}{\gamma_3 \cdot \varepsilon_4}\right)$ are the phase shifts at the top and bottom interfaces of the PMMA and the bottom interface of silver. $\gamma_1 = \sqrt{\beta^2 - k_0^2 n_1^2}$, $\kappa_2 = \sqrt{k_0^2 n_2^2 - \beta^2}$, $\gamma_3 = \sqrt{\beta^2 - k_0^2 n_3^2}$, $\kappa_4 = \sqrt{k_0^2 n_4^2 - \beta^2}$ are the wave vectors normal to the propagation plane in the air ($n_1$), PMMA layer ($n_2$), silver film ($n_3$), and glass substrate ($n_4$) correspondingly. $k_0$ is the wave vector in vacuum. $\beta$ is the effective propagation constant of waveguide mode and $n_{eff} = \beta / k_0$ is the mode effective index. d and t are the thicknesses of PMMA and silver. In our experiment, the thickness of silver is 50 nm. The refractive indices of PMMA, silver, and quartz at 405 nm are 1.49, 0.173, and 1.55, respectively.

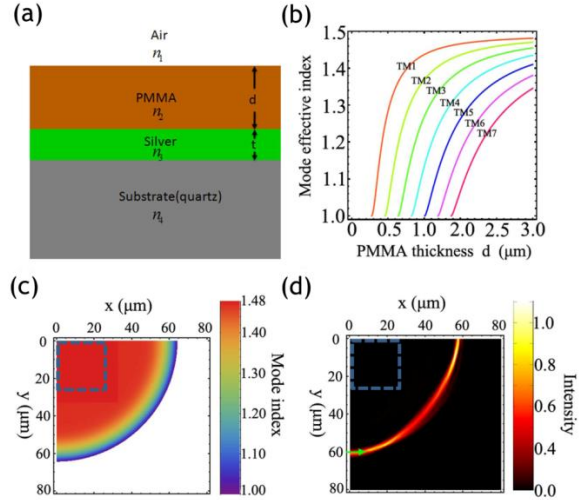

Fig. 4. (a) Schematic picture of waveguide; (b) thickness dependence of effective mode index; (c) effective mode index profile of the structure; (d) numerical simulation results of light propagating around the corner.

Figure. 4(b) shows the thickness dependence of the effective index of the TM mode. As the thickness of the PMMA layer of the coupling region is around 500 nm in our structure, most of the incident light is coupled into TM1 mode. A photon launched into the *l*-th mode of the waveguide stays in this mode as along as thickness changes gradually, so we can transfer the surface profile (Fig. 3(d)) to effective index contour based on the thickness dependence of the TM1 mode (Fig. 4(c)). The mode index increases from 1.0 to 1.48 gradually as r decreases from 65 μm to 50 μm, where r is the distance from the top left corner. Therefore, a graded PMMA waveguide surrounding the corner is formed and its effective index is determined by gradient thickness.

A two-dimensional full wave simulation (COMSOL Multiphysics) is performed to simulate the propagation of light around the corner in the graded waveguide. The resulting effective index profile is given as Fig. 4(c). A Gaussian beam waist of 4 μm on the left is input into the waveguide with the wavelength at 405 nm. Fig. 4(d) is the simulated trajectory of this laser beam in the graded waveguide. At the entrance, the incident beam propagates into the region along the +x direction. After penetrating into the graded index region, the beam's path gradually bends toward the corner, where the PMMA layer is thicker and the effective index is larger. In the process, a continuous curve is formed by the light beam around the corner. Lastly, the light propagates upward though the exit. The beam can be transported around the corner along a continuous curve without any reflection or scattering loss. This result shows that our structure can perform well as an optical bending device.

## 4. Optical measurement results

In the experiment, a 405 nm laser is coupled into the waveguide through a grating on the surface of silver film. The grating consists of 13 grooves of 130 nm, with depth of 20 nm and period of 310 nm. CdSe/ZnS quantum dots are doped into the PMMA layer, which could be excited by a 405 nm laser and has a fluorescence peak at 605 nm. The fluorescence emission from the quantum dots is collected using a 50×/0.8 Zeiss microscope objective and imaged on a charge-coupled device camera. After filtering the 405 nm laser using a long wave pass filter, the propagation trajectory of the 405 nm laser inside the waveguide could be visualized using the fluorescence image. A weak white light source (halogen lamp) is used to illuminate the structure on the backside, which gives the fluorescence image a red background.

Figure. 5 shows the fluorescence images of the 405nm laser propagation in the structures. In Fig. 5(a), we focus the 405 nm laser on a grating clinging to the boundary far away from the sharp corner. Light propagates along the boundary. Due to the sloping shape of the PMMA layer against the side of the top wafer (Fig. 2(d)), the effective index is increased as the boundary is approached. A graded waveguide is thus formed, and light can be easily confined and transported in it.

In Fig. 5(b), the laser is focused on the grating near the corner. If we look at the picture closely(a close-up is shown), three optical beams are excited, which correspond to three waveguide modes. Two of them, which correspond to the surface plasmon mode and TM2 mode, attenuate quickly and cannot propagate long enough to reach around the corner. The middle one, which corresponds to TM1 mode, bends around the sharp corner smoothly with low reflection loss. This picture agrees well with the simulated results given in Fig. 4(d).

Besides this bending effect, our structure demonstrates some other interesting properties. In Fig. 5(c), the light is coupled into the waveguide at a distance from the boundary. We can observe two waveguide modes being excited. Due to the sloping shape of the waveguide, the effective index increases as the boundary nears. As a result, the optical beam bends toward the boundary. The dispersion is different for different modes, so we can see two separate trajectories, as shown in Fig. 5(c).

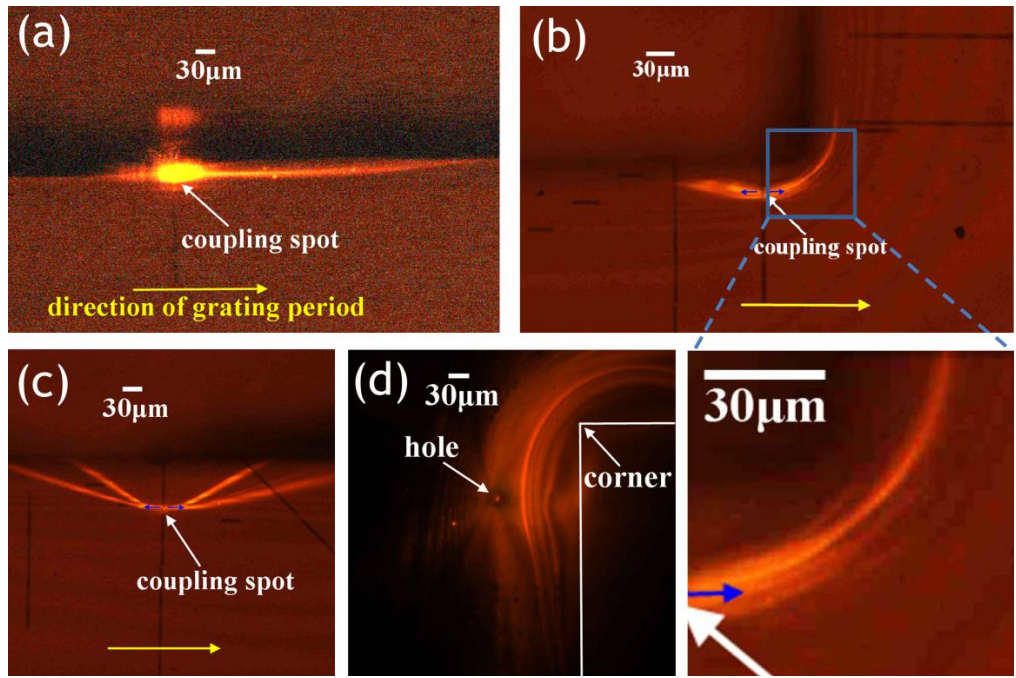

Fig. 5. Fluorescence image of light propagation in the waveguide. (a) light propagation along the boundary of waveguide; (b) light bending around the sharp corner; (c) mode splitting in thick waveguide; (d) radiation of a point source around the sharp corner; (e) magnified bending area taken from the frame in (b).

In the above experiments, the light is coupled into the waveguide through grating. The propagation direction can be controlled and is always perpendicular to the grating. The directed light beams are then excited in Figs. 5(a to c). However, in Fig. 5(d) we instead drill a hole with a diameter of 2 μm through the silver film. As the laser illuminates the sample from the backside, this hole can work as a point source to couple light into the waveguide. In Fig. 5(d), the hole emits light in all directions. However, due to the distribution of the effective index around the hole, the omnidirectional radiation is confined within the waveguide. The radiation energy is still transported around the corner.

**5. Discussion and outlook**

In the previously discussed experiments, we only use a 405 nm laser to excite the waveguide modes. However, our proposed system is not limited to this specific wavelength. In principle, our structures can be used as optical bending devices for all visible light. According to the theory of optical waveguides, the cut-off wavelength is determined by the waveguide's thickness, and all light with wavelengths below the cut-off wavelength can be transported along the waveguide. We can therefore enlarge the working bandwidth by increasing the thickness of the PMMA. This thickness profile can be adjusted to over 2.0 μm during the spin coating process by changing the volume ratio of PMMA to solvent, or by changing the spin speed. The thickness is determined by the natural surface tension of PMMA, So the distribution of effective index is completely continuous without any sudden changes, resulting in very small scattering loss. This result gives our structures a major advantage over other inhomogeneous systems.

## 6. Conclusion

In summary, we fabricate a broadband transformation device using the spin-coating technique. A gradient-thickness optical waveguide around a 90 ° corner is obtained using this method. Fluorescence imaging is used to trace the trajectory of light propagation along the waveguide. The measured results show that a blue light can be transported around the corner continuously without reflection. The ohmic loss and scattering loss can be very minimal. The behavior of the waveguide is confirmed with numerical simulations. The fabrication method reported in this work is simple, inexpensive, and easily controlled. These attributes are useful for application in many other transformation devices in the future.

**Acknowledgement**

This work was financially supported by the National Natural Science Foundation of China (No. 11021403, 11074119, 60990320 and 11004102), and by the National Key Projects for Basic Researches of China (No. 2010CB630703, 2012CB921500 and 2012CB933501).